\documentstyle[prd,aps]{revtex}

\input epsf
\epsfverbosetrue

\def\Journal#1#2#3#4{{#1} {\bf #2} (#4) #3}
\def\MPL{Mod. Phys. Lett. A}

\def\NPB{Nucl. Phys. B}
\def\NPBOLD{Nucl. Phys.}
\def\NPSUPPL{Nucl. Phys. Proc. Suppl.}
\def\PREPO{Phys. Rep.}
\def\PLB{Phys. Lett. B}
\def\PLBOLD{Phys. Lett.}
\def\PRL{Phys. Rev. Lett.}
\def\RMP{Rev. Mod. Phys.}
\def\PRD{Phys. Rev. D}

\def\PTP{Prog. Theor. Phys.}
\def\JHEP{JHEP}
\def\EPJ{Euro. Phys. J. C}
\def\JETPUSSR{JETP (USSR)}
\def\ZETP{Zh. Eksp. Teor. Piz.}

\def\mapgeq{\mathbin{\lower.3ex\hbox{$\buildrel>\over{\smash{\scriptstyle\sim}\vphantom{_x}}$}}}
\def\mapleq{\mathbin{\lower.3ex\hbox{$\buildrel<\over{\smash{\scriptstyle\sim}\vphantom{_x}}$}}}
\def\mapgeqeq{\mathbin{\lower.3ex\hbox{$\buildrel>\over{\smash{\scriptstyle\approx}\vphantom{_2}}$}}}
\def\mapleqeq{\mathbin{\lower.3ex\hbox{$\buildrel<\over{\smash{\scriptstyle\approx}\vphantom{_2}}$}}}

\tighten
\draft
\begin{document} 
\bibliographystyle{prsty}

\title{
Large Mixing Angle MSW Solution in an $SU(3)_L \times U(1)_N$ Gauge Model\\
with Two-loop Radiative Mechanism
}

\author{
Teruyuki Kitabayashi$^a$
\footnote{E-mail:teruyuki@post.kek.jp}
and Masaki Yasu${\grave {\rm e}}^b$
\footnote{E-mail:yasue@keyaki.cc.u-tokai.ac.jp}
}

\address{\vspace{5mm}$^a$
{\sl Accelerator Engineering Center} \\
{\sl Mitsubishi Electric System \& Service Engineering Co.Ltd.} \\
{\sl 2-8-8 Umezono, Tsukuba, Ibaraki 305-0045, Japan}
}
\address{\vspace{2mm}$^b$
{\sl Department of Natural Science\\School of Marine
Science and Technology, Tokai University}\\
{\sl 3-20-1 Orido, Shimizu, Shizuoka 424-8610, Japan\\and\\}
{\sl Department of Physics, Tokai University} \\
{\sl 1117 KitaKaname, Hiratsuka, Kanagawa 259-1292, Japan}}
\date{TOKAI-HEP/TH-0102, February, 2001}
\maketitle

\begin{abstract}
An $SU(3)_L \times U(1)_N$ gauge model, where $E^{-i}$ ($i$=1,2,3) as heavy leptons are placed in the third member of lepton triplets, is shown to provide the large mixing angle (LMA) MSW solution to the solar neutrino problem. By introducing approximate conservation of $L_e-L_\mu-L_\tau$ ($\equiv L^\prime$) into the model, we have estimated $\Delta m_\odot^2$ $\sim$ $\sqrt{2}\epsilon \Delta m_{atm}^2m_{E^1}$/$(m_{E^2}+m_{E^3})$, where $\epsilon$ represents $L^\prime$-breaking effects and $m_{E^i}$ is a mass of $E^{-i}$. Since $\Delta m_{atm}^2 \sim 3 \times 10^{-3}$ eV$^2$, the LMA solution is obtained for $\epsilon\sim 0.01-0.1$ in the case of $m_{E^1}$ $\sim$ $m_{E^{2,3}}$.  This mechanism utilizes two-loop radiative mechanism, which contains a triplet version of the singly charged Zee boson, $h^+$, given by $\xi$ = ($\xi^{++}$, ${\bar \xi}^+$, $h^+$)$^T$.  Almost bimaximal mixing is dynamically realized by the approximate equality between the $\xi$-couplings of the first family to the second and third families.
\end{abstract}
\vspace{2mm}
The recent analysis of solar neutrino oscillation data done by the Super-Kamiokande collaboration has so far shown no evidence for a distortion of the energy spectrum, nor for Earth regeneration effects, nor for seasonal variations of the neutrino flux \cite{RecentSK}. These observations suggest that solutions with large mixing angles are favored while solutions with small mixing angles as well as the vacuum oscillation (VO) solution are disfavored at the 95\% confidence level.
\footnote{However, see also other analyses in Ref.\cite{SolarComment} for saving the disfavored solutions.}  Solar neutrino oscillations \cite{Solar} characterized by large mixing angles are controlled by $\Delta m_\odot^2 \sim 10^{-5}-10^{-4}$ eV$^2$ for the large mixing angle (LMA) MSW solution, $\Delta m_\odot^2 \sim 10^{-7}$ eV$^2$ for the LOW solution with low mass and low probability,  $\Delta m_\odot^2 \sim 10^{-9}$ eV$^2$ for the quasi-VO (QVO) solution similar to VO but with small matter corrections taken into account \cite{QVO} and $\Delta m_\odot^2 \sim 10^{-10}$ eV$^2$ for the VO solution.  Such neutrino oscillations with large mixing angles are also observed for atmospheric neutrinos \cite{Kamiokande}.  The preliminary analysis on long-baseline neutrino oscillations done by K2K \cite{RecentK2K} has shown that $\Delta m_{atm}^2 \approx 3 \times 10^{-3}$ eV$^2$ well reproduces their data.  Therefore, there are two mass scales, $\Delta m_{atm}^2={\mathcal O}$($10^{-3}$) eV$^2$ and $\Delta m_\odot^2 \le {\mathcal O}$($10^{-4}$) eV$^2$, which indicate the hierarchy of $\Delta m_{atm}^2\gg\Delta m_\odot^2$.  Neutrino masses \cite{EarlyMassive} are implied to be about several $\times$ 10$^{-2}$ eV and such tiny neutrino masses can be generated by the seesaw mechanism or the radiative mechanism \cite{SeeSaw,1-loop,2-loop}. 

Since both oscillations are now characterized by large mixing angles, the observed pattern of the neutrino oscillations is compatible with bimaximal mixing scheme proposed earlier \cite{Mixing,NearlyBiMaximal}.  One of the underlying physics behind bimaximal mixing is the presence of a new symmetry based on a lepton number of $L_e - L_\mu - L_\tau$ ($\equiv$ $L^\prime$) \cite{Lprime,EarlierLprime}.  Especially, in radiative mechanism of creating tiny neutrino masses \cite{1-loop,2-loop,Radiative,ZeeType}, the use of $U(1)_{L^\prime}$ based on $L^\prime$ has recently been advocated \cite{1-loopLprime,2-loopLprime}.  The $U(1)_{L^\prime}$ symmetry applied to three active Majorana neutrino mass terms guarantees the hierarchy of $\Delta m_{atm}^2\gg\Delta m_\odot^2$ with the maximal mixing for solar neutrinos.  However, $\Delta m_\odot^2$ = 0 is also derived and is experimentally unacceptable.  Finite but tiny amount of $\Delta m_\odot^2$ can be induced if tiny $U(1)_{L^\prime}$-breaking interactions are present.  There is a scenario that naturally contains tiny $U(1)_{L^\prime}$-breaking interactions in the radiative mechanism, where one-loop radiative mechanism is responsible for $\Delta m_{atm}^2$ while two-loop radiative mechanism accompanied by $U(1)_{L\prime}$-breaking effects creates $\Delta m_\odot^2$ \cite{1loop2loop}.  Since two-loop radiative effects are much more suppressed than one-loop radiative ones, it is naturally expected to obtain $\Delta m_{atm}^2\gg\Delta m_\odot^2$.

In the Zee's radiative mechanism \cite{1-loop}, a singly charged $SU(2)_L$-singlet Higgs scalar, $h^+$, which couples to charged lepton - neutrino pairs, is a key ingredient.  It has been claimed that, in $SU(3)_L \times U(1)_N$ gauge models \cite{SU3U1,SU3U1Topics}, $h^+$ together with the standard Higgs, $\phi$, can be unified into a Higgs triplet either with ($\phi^+$, $\phi^0$, $h^+$) or with ($\phi^0$, $\phi^-$, $h^+$) and that $\Delta m_{atm}^2$ from one-loop radiative mechanism \cite{Okamoto} as well as $\Delta m_\odot^2$ from two-loop radiative mechanism \cite{Kita00} are successfully generated.  However, we have ended up with the VO and QVO solutions but not with the most favorable LMA solution \cite{Kita00}.  The reason lies in the fact that two-loop effects are too suppressed to obtain $\Delta m_\odot^2/\Delta m_{atm}^2={\mathcal O}$(10$^{-2}$) for the LMA solution.  
Roughly speaking, we estimated $m^{(1)}$ $\mapgeq$ $fm^2_\tau/(16\pi^2v)$ as a one-loop mass scale and $m^{(2)}$ $\mapleq$ $\epsilon f^2v/(16\pi^2)^2$ as a two-loop mass scale, where $f$ represents an $h^+$-coupling to a neutrino-charged lepton (or neutrino-heavy lepton) pair, $v$ $\sim$ $v_{weak}$ = $( 2{\sqrt 2}G_F)^{-1/2}$ (= 174 GeV) and $\epsilon$, which is presumably much smaller than 1, measures $L^\prime$-breaking effects.  This estimate gives $\Delta m_{atm}^2$ $\sim$ $m^{(1)2}$ and $\Delta m_\odot^2$ $\sim$ $m^{(1)}m^{(2)}$ because of $m^{(1)}$ $\gg$ $m^{(2)}$, leading to $\Delta m_\odot^2$/$\Delta m_{atm}^2$ $\sim$ $m^{(2)}$/$m^{(1)}$ $\mapleq$ $\epsilon f/16\pi^2\times 10^4$ ($<$ $f/16\pi^2\times 10^4$), which must include the magnitude of ${\mathcal O}$(10$^{-2}$) for the LMA solution.  Therefore, the LMA solution at least requires $f$ $\mapgeq$ $10^{-4}$, which is, however, not the case because $\Delta m_{atm}^2$ $\sim$ $3\times 10^{-3}$ eV$^2$ fixes $f$ to be about 10$^{-7}$, thereby, leading to $\Delta m_\odot^2$ $\mapleq $10$^{-8}$ eV$^2$, which covers the VO and QVO solutions as mentioned. 

In the present article, we seek other possibilities to be consistent with the LMA solution within the $SU(3)_L \times U(1)_N$ frameworks.  A model with heavy leptons with the same charges as those of charged leptons \cite{HeavyE} turns out to be relevant to provide the LMA solution.  Since lepton triplets consist of ($\nu^i_L$, $\ell^{-i}_L$, $E^{-i}_L$) with $i$ = 1, 2, 3, this charge assignment requires the scalar of $h^+$ together with a new $SU(2)_L$-doublet Higgs ($\xi^{++}$, ${\bar \xi}^+$) to be included in a Higgs triplet of ($\xi^{++}$, ${\bar \xi}^+$, $h^+$).  It is demonstrated that all neutrino masses are generated by two-loop radiative mechanism but not by one-loop radiative mechanism \cite{2-loopLprime}. The suppression of $\Delta m_\odot^2$ is essentially ascribed to tiny breaking of $U(1)_{L^\prime}$ without the extra loop-suppression.  The model is found to yield $\Delta m_\odot^2$ $\sim$ $\sqrt{2}\epsilon m_{E^1}$/$(m_{E^2}+m_{E^3})$ $\Delta m_{atm}^2$, where $m_{E^i}$ is a mass of the $i$-th heavy lepton, which reproduces the LMA solution to solar neutrino problem for $\epsilon\sim (0.01-0.1)$ with $m_{E^1}\sim m_{E^{2,3}}$.

In $SU(3)_L \times U(1)_N$ models, the minimal set of Higgs scalars consists of three triplets to be denoted by $\eta$, $\rho$ and $\chi$ in Eq.(\ref{Eq:Higgs}), which are minimal because, without these scalars, masses of triplet quarks would not be generated.  In addition, our model uses a triplet version of the Zee scalar \cite{1-loop}, $\xi$, and a doubly charged scalar of the Zee-Babu type \cite{2-loop}, $k^{++}$, that initiates two-loop radiative mechanism. The particle content is specified by the $U(1)_N$ quantum number, $N/2$, related to the hypercharge, $Y$, as $Y$=$\lambda^8/\sqrt{3}+N$, leading to $Q_{em}=(\lambda^3+Y)/2$, where $\lambda^a$ is the $SU(3)$ generator with Tr$(\lambda^a \lambda^b)=2\delta^{ab}$ ($a,b$=1$\sim$8).  Summarized as follows is the whole particle content in our model, where the quantum numbers are specified by ($SU(3)_L$, $U(1)_N$):
\begin{eqnarray}
&&\psi^{i=1,2,3}_L
     =\left(\nu^i,\ell^i,E^i\right)_L^T:\left(\textbf{3},-2/3\right),\quad
\ell^{1,2,3}_R                         :\left(\textbf{1},-1 \right),\quad  
E^{1,2,3}_R                            :\left(\textbf{1},-1 \right),
\end{eqnarray}
\label{Eq:Leptons}
as leptons, where we have denoted heavy leptons $E^{-i}$ by $E^i$ ($i$=1, 2, 3),
\begin{eqnarray}
&&Q^{1}_L=\left(u^1,d^1,d^{\prime 1} \right)_L^T
	   :\left(\textbf{3},0 \right),\quad
Q^{i=2,3}_L=\left(d^i,-u^i,u^{\prime i}\right)_L^T
       :\left(\textbf{3}^\ast,1/3\right),
\nonumber \\
&&u^{1,2,3}_R      : \left( \textbf{1}, 2/3 \right),\quad 
d^{1,2,3}_R      : \left( \textbf{1},-1/3 \right),\quad 
u^{\prime 2,3}_R : \left( \textbf{1}, 2/3 \right),\quad
d^{\prime 1  }_R : \left( \textbf{1},-1/3 \right),
\end{eqnarray}
\label{Eq:Quarks}
as quarks, and
\begin{eqnarray}
&& \eta = \left(\eta^0,\eta^-,\overline{\eta}^-\right)^T
	                   : \left( \textbf{3}, -2/3 \right), \quad
    \rho = 
       \left(\rho^+,\rho^0,\overline{\rho}^0\right)^T
	   :\left( \textbf{3}, 1/3 \right),
\nonumber \\
&&  \chi = 
       \left(\chi^+,\overline{\chi}^0,\chi^0 \right)^T
	   :\left( \textbf{3}, 1/3 \right), \quad  
	\xi = 
       \left(\xi^{++},\overline{\xi}^+,\xi^+\right)^T
	   :\left( \textbf{3}, 4/3 \right),
\label{Eq:Higgs}
\end{eqnarray}
as triplet Higgs scalars and
\begin{equation}
k^{++}:(\textbf{1},2),
\label{Eq:k++}
\end{equation}
as a singlet Higgs scalar.  The Higgs scalars have the following vacuum expectation values (VEV's):
\begin{equation}
\langle 0 \vert\eta\vert 0 \rangle  = \left(v_\eta,0,      0      \right)^T, \quad 
\langle 0 \vert\rho\vert 0 \rangle  = \left(0,     v_\rho, 0      \right)^T, \quad 
\langle 0 \vert\chi\vert 0 \rangle  = \left(0,     0,      v_\chi \right)^T,
\label{Eq:VEV}
\end{equation}
and quarks and leptons will acquire masses via these VEV's, where the orthogonal choice of these VEV's is ensured by appropriate Higgs interactions to be shown in Eq.(\ref{Eq:Orthgonal}). Quarks and leptons are collaborating to cancel gauge anomalies \cite{HeavyE}, where the obvious cancellation of the pure $SU(3)_L$ anomaly is done by two triplets and two anti-triplets.

The lepton number, $L$, associated with $\psi^i_L$ is assigned to other particles such that their possible interactions, which are $SU(3)_L\times U(1)_N$-singlets, maximally conserve the lepton number.  For our newly introduced Higgs scalars, the possible interactions are $\psi^i_L\psi^j_L\xi$, $\ell^i_R\ell^j_Rk^{++}$ and $E^i_RE^j_Rk^{++}$, which can be used to determine $L$ = $-$2.  Other $L$-assignment is shown in Table \ref{Tab:Lnumber}.  As a result, all Yukawa interactions can be described by the $L$-conserving ones.

In order to explain the observed behavior in neutrino oscillations, we also introduce another lepton number $L_e-L_\mu-L_\tau$ (=$L^\prime$) and require its (approximate) conservation in our interactions. It is the tiny $L^\prime$-breaking interactions that lead to the hierarchy of $\Delta m_{atm}^2\gg\Delta m_\odot^2$. In order to import $L^\prime$-breaking interactions, we employ the duplicate of $k^{++}$:
\begin{equation}
	   k^{\prime ++}: (\textbf{1},2),
\label{Eq:kprime++}
\end{equation}
but with the different $L'$ of $L'$ = $-$2 so as to develop possible Yukawa interactions, which will respect all relevant symmetries.  The $L^\prime$-assignment to other participating particles can been read off from Table \ref{Tab:Lnumber}. 

In addition to these global symmetries, another discrete symmetry based on $Z_2$ is used to suppress  dangerous flavor-changing-neutral-currents (FCNC) interactions.  Since our model contains quarks with the same charge, whose mass terms can be generated by $\rho$ and $\chi$ between $Q_L^1$ and down-type quarks and by $\rho^\dagger$ and $\chi^\dagger$ between $Q_L^{2,3}$ and up-type quarks, FCNC is induced at the phenomenologically unacceptable level \cite{FCNCSU3}.  To avoid these interactions,  Yukawa interactions must be constrained such that a triplet quark flavor gains a mass from only one Higgs field \cite{FCNC}.  The lepton sector also contains the similar FCNC problem because $\ell^i$ and $E^i$ ($i$=1, 2, 3) has the same charge.  It turns out that such a constraint on FCNC is satisfied by introducing the following $Z_2$ symmetry into the model:
\begin{eqnarray}
&& \psi_L^{1,2,3}   \rightarrow \psi_L^{1,2,3},     \quad
   \ell_R^{1,2,3}   \rightarrow \ell_R^{1,2,3},     \quad
   E_R^{1,2,3}      \rightarrow -E_R^{1,2,3},
\nonumber \\
&& Q_L^{1,2,3}      \rightarrow    Q_L^{1,2,3},      \quad
   u_R^{1,2,3}      \rightarrow    u_R^{1,2,3},      \quad
   d_R^{1,2,3}      \rightarrow    d_R^{1,2,3},      \quad
   u_R^{\prime 2,3} \rightarrow   -u_R^{\prime 2,3}, \quad
   d_R^{\prime 1}   \rightarrow   -d_R^{\prime 1},
\nonumber \\
&& \eta             \rightarrow    \eta,             \quad
   \rho             \rightarrow    \rho,             \quad
   \chi             \rightarrow   -\chi,             \quad
   \xi              \rightarrow    \xi,              \quad
   k^{++}           \rightarrow    k^{++},           \quad
   k^{\prime ++}    \rightarrow    k^{\prime ++}.
\end{eqnarray}

Our Higgs interactions are, now, given by self-Hermitian terms composed of $\phi_\alpha\phi^\dagger_\beta$ ($\phi$ = $\rho$, $\eta$, $\chi$, $\xi$, $k^{++}$, $k^{\prime ++}$), which include the potential term of $V_{\eta\rho\chi}$ 
\begin{eqnarray}\label{Eq:Orthgonal}
&V_{\eta\rho\chi} = \lambda_1 \vert \eta \times \rho \vert^2
      + \lambda_2 \vert\rho \times \chi\vert^2
      + \lambda_3 \vert\chi \times \eta\vert^2
\end{eqnarray}
with the definition of $a \times b$ $\equiv$ $\epsilon^{\alpha\beta\gamma}a_\beta b_\gamma$ and by non self-Hermitian Higgs potentials, $V_0$ and $V_b$:
\begin{eqnarray}
V_0 &=& \lambda_4 (\eta^\dagger \chi)(\xi^\dagger \chi)
      + \lambda_5 (\eta^\dagger \rho)(\xi^\dagger \rho)
      + \mu_0 \xi^\dagger \eta k^{++} 
	  + (h.c.),
\nonumber \\
V_b &=& \mu_b \xi^\dagger \eta k^{\prime ++} + (h.c.),
\label{Eq:HiggsPotential}
\end{eqnarray}
where $\lambda_i$ ($i=1\sim 5$) stands for a coupling constant and $\mu_0$ and $\mu_b$, respectively, denote $L^\prime$-conserving and $L^\prime$-breaking mass scale. The Higgs potential of $V_0$ conserves $L^\prime$ and is responsible for the emergence of the bimaximal mixing while $V_b$ breaks the $L^\prime$-conservation and is responsible for the emergence of the hierarchy of $\Delta m^2_{atm} \gg \Delta m^2_\odot$.  The Yukawa interactions are given by the following lagrangian:
\begin{eqnarray}
-{\mathcal L}_Y &=& 
     \sum_{i=2,3}f_{[1i]}\epsilon^{\alpha\beta\gamma}
     \overline{\left(\psi_{\alpha L}^1 \right)^c} \psi_{\beta L}^i \xi_\gamma
   + \sum_{i=1,2,3} \overline{\psi_L^i}
     \left( f_\ell^i \rho \ell_R^i + f_E^i \chi E_R^i \right)
\nonumber \\
&& + \sum_{i=2,3} 
         \left[
             f^\ell_{\{i1\}}\overline{(\ell_R^i)^c}\ell_R^1 + f^E_{\{i1\}}\overline{(E_R^i)^c}E_R^1 
		 \right]k^{++} 
   + \frac{1}{2} 
         \left[
	         f^{\prime\ell}_{\{11\}}\overline{(\ell_R^1)^c}\ell_R^1 + f^{\prime E}_{\{11\}}\overline{(E_R^1)^c}E_R^1
         \right]k^{\prime ++} 
\nonumber \\
&& + \overline{Q_L^1}\left( \eta U_R^1 + \rho D_R^1 + \chi D_R^{\prime 1} \right)
   + \sum_{i=2,3} \overline{Q_L^i}
     \left( \eta^\ast D_R^i + \rho^\ast U_R^i + \chi^\ast U_R^{\prime i} \right)
   + (h.c.),
\label{Eq:Yukawa}
\end{eqnarray}
where $f$'s are Yukawa couplings with $f_{[ij]}=-f_{[ji]}$ and $f^{\ell,E}_{\{ij\}}=f^{\ell,E}_{\{ji\}}$, the lepton mass terms are assumed to be diagonal for simplicity and right-handed quarks are denoted by $U_R^i=\sum_{j=1}^3 f_{uj}^i u_R^j$, $D_R^i=\sum_{j=1}^3 f_{dj}^i d_R^j$, $U_R^{\prime i}=f_{u^\prime 2,3}^i u_R^{\prime 2,3}$ and $D_R^{\prime 1}=f_{d^\prime 1}^i d_R^{\prime 1}$.

We here note four comments on physical aspects of our interactions:
\begin{enumerate}
\item the suppression of FCNC is ensured by the absence of $\chi D^1_R$ and $\rho D^{\prime 1}_R$ for $Q^1_L$, of $\chi^\ast U^i_R$ and $\rho^\ast U^{\prime i}_R$ for $Q^{2,3}_L$ and of $\chi \ell^i_R$ and $\rho E^i_R$ for $\psi^i_L$;
\item the orthogonal choice of VEV's of $\eta$, $\rho$ and $\chi$ as in Eq.(\ref{Eq:VEV}) is supported by $V_{\eta\rho\chi}$ if all $\lambda$'s are negative.  It is because $V_{\eta\rho\chi}$ gets lowered if $\eta$, $\rho$ and $\chi$ develop VEV's.  So, one can choose VEV's such that $\langle 0\vert\eta_1 \vert 0\rangle$ $\neq$ 0, $\langle 0\vert\rho_2 \vert 0\rangle$ $\neq$ 0 and $\langle 0\vert|\chi_3 \vert 0\rangle$ $\neq$ 0;
\item the breaking of $L$-conservation is supplied by $(\eta^\dagger\chi)(\xi^\dagger\chi)$ and $(\eta^\dagger\rho)(\xi^\dagger\rho)$ in $V_0$; and
\item the breaking of $L^\prime$-conservation is supplied by $\xi^\dagger\chi k^{\prime ++}$ in $V_b$.
\end{enumerate}
The additional effects of applying the $Z_2$ symmetry to the model include the absence of interactions given by
\begin{enumerate}
\item $(\ell^1_RE^i_R+\ell^i_RE^1_R)k^{++}$ ($i$=2,3) and $\ell^1_RE^1_Rk^{\prime ++}$ that would give extra contributions to neutrino masses,
\item $\eta\rho\chi$ that would enhance the orthogonal choice of VEV's, and
\item $(\eta^\dagger \rho)(\xi^\dagger \chi)$ that would induce one-loop radiative interactions.
\end{enumerate}
Therefore, the suppression of FCNC simultaneously allows us to formulate our $SU(3)_L \times U(1)_N$ model with two-loop radiative mechanism but without one-loop radiative mechanism.  The absence of one-loop radiative mechanism is essential for our discussions to obtain the LMA solution to the solar neutrino problem.

Now, let us explain how our $SU(3)_L \times U(1)_N$ model yields the LMA solution to the solar neutrino problem.  Tiny neutrino masses are generated by two-loop radiative mechanism, where the interplay between $\xi$ and $k^{++}$ provides mixing masses for $\nu_{eL}$ and $\nu_{\mu L, \tau L}$ while the one between $\xi$ and $k^{\prime ++}$ provides tiny mixing masses among $\nu_{\mu L,\tau L}$.  Indeed, these mixings are generated by interactions corresponding to $L^\prime$-conserving diagrams as in Fig.\ref{Fig:loopDiagrams}(a) and to $L^\prime$-violating diagrams as in Fig.\ref{Fig:loopDiagrams}(b), which contain Higgs couplings described by two types of non self-Hermitian Higgs potentials in Eq.(\ref{Eq:HiggsPotential}).  These interactions can be used to derive $\Delta m_\odot^2$ $\sim$ $\sqrt{2}\epsilon m_{E^1}$/$(m_{E^2}+m_{E^3})$ $\Delta m_{atm}^2$, where $\epsilon$ represents $L^\prime$-breaking effects and $m_{E^i}$ is the mass of $E^i$. The LMA solution will correspond to $\epsilon\sim 0.01-0.1$ for $m_{E^1}$ $\sim$ $m_{E^{2,3}}$.

From two-loop calculations, we obtain the following neutrino mass matrix parameterized by 
$m_{1i}$ calculated from $L^\prime$-conserving two-loop diagrams and by $m_{ij}^\prime $ calculated from $L^\prime$-violating diagrams ($i,j$=2,3):
\begin{equation}
M_\nu =   
    \left(
    \begin{array}{ccc}
         0            & m_{12}        & m_{13} \\
         m_{12} & m_{22}^\prime  & m_{23}^\prime  \\
         m_{13} & m_{23}^\prime  & m_{33}^\prime  \\
    \end{array}
    \right),
\label{Eq:MassMatrix}
\end{equation}
where
\begin{eqnarray}
&&m_{1i} = -2f_{[1i]}c_0\mu_0, \quad
m^\prime_{ij} = -f_{[1i]}f_{[1j]} c_b\mu_b 
\label{Eq:MajoranaMass1}
\end{eqnarray}
with
\begin{eqnarray}
c_0 &=& \sum_{i=2,3}f_{[1i]}
	\left[ \lambda_4 f^\ell_{\{i1\}}m_{\ell^i}m_e  v_\chi^2 
	       I(m_{\ell^i}^2, m_e^2, m_{\xi^+}^2, m_{\overline{\eta}^-}^2, m_{\xi^+}^2, m_k^2) 
	\right. 
\nonumber \\
 && \left. + \lambda_5 f^E_{\{i1\}}m_{E^i}m_{E^1}v_\rho^2 
	         I(m_{E^i}^2,m_{E^1}^2,m_{\overline{\xi}^+}^2,m_{\eta^-}^2,m_{\overline{\xi}^+}^2,m_k^2)
    \right],
\nonumber \\
c_b  &=& \lambda_4f^{\prime\ell}_{\{11\}} m_e^2 v_\chi^2 
        I(m_e^2, m_e^2,m_{\xi^+}^2,m_{\overline{\eta}^-}^2,m_{\xi^+}^2,m_{k^\prime}^2)
\nonumber \\
  && + \lambda_5f^{\prime E}_{\{11\}} m_{E^1}^2 v_\rho^2
	    I(m_{E^1}^2, m_{E^1}^2,m_{\overline{\xi}^+}^2,m_{\eta^-}^2,m_{\overline{\xi}^+}^2,m_{k^\prime}^2).
\label{Eq:MajoranaMass2}
\end{eqnarray}
Here, the mass of the $i$-th charged lepton and the mass of the $i$-th heavy lepton are, respectively, denoted by $m_{\ell i}$ ($\equiv f_{\ell i}^i v_\rho$) and $m_{E^i}$ ($\equiv f_{E^i}^i v_\chi$) and masses of Higgs scalars are denoted by the subscripts in terms of their fields
. From the Appendix of Ref.\cite{Kita00}, we obtain the 2-loop integral, $I$, estimated as follows:
\begin{eqnarray}
I(m_c^2,m_d^2.m_1^2,m_2^2,m_3^2,m_4^2)
    &=&\frac{J(m_c^2,m_d^2,m_1^2,m_2^2,m_4^2)-J(m_c^2,m_d^2,m_1^2,m_3^2,m_4^2)}{m_2^2-m_3^2},
\label{Eq:I_1} \\
J(m_c^2,m_d^2,m_1^2,m^2,m_4^2)
    &=&\frac{1}{m_4^2} G(m_c^2,m_1^2,m_4^2)G(m_d^2,m^2,m_4^2),
\label{Eq:I_2} \\
G(x,y,z) &=& \frac{1}{16\pi^2} \frac{x \ln (x/z) - y \ln (y/z)}{x-y}.
\label{Eq:I_3}
\end{eqnarray}
The integral of Eq.(\ref{Eq:I_1}) can be approximated by the chief contributions from $m^2_{k,k^\prime}$ in Eq.(\ref{Eq:I_3}) for $m^2_{k,k^\prime}\gg$ (other masses squared).  Therefore, it can be taken to be universal for $c_0$ and $c_b$ and is to be denoted by $I_{univ}$.  Hereafter, the dominance of $m^2_{k,k^\prime}$ in masses of the model is assumed for the following analyses.  We also assume that $\lambda_4$ $\sim$ $\lambda_5$.

The form of $M_\nu$ with the anticipation of $m_{1i} \gg m_{ij}$ gives the following expressions of $\Delta m_{atm}^2$ and $\Delta m_\odot^2$:
\begin{eqnarray}
&& \Delta m_{atm}^2 = m_{12}^{2}+m_{13}^{2} (\equiv m_\nu^2),
\nonumber \\
&& \Delta m_\odot^2 = 2 
	\left(  m_{22}^\prime  \cos^2\vartheta
		+ 2 m_{23}^\prime  \sin\vartheta\cos\vartheta 
		+   m_{33}^\prime  \sin^2\vartheta
	\right)m_\nu,
\label{Eq:DeltaM}
\end{eqnarray}
where $\vartheta$ is the mixing angle for solar neutrinos and is defined by $\cos\vartheta = m_{12}/m_\nu$ and $\sin\vartheta = m_{13}/m_\nu$.  The maximal atmospheric neutrino mixing can be achieved by setting $m_{12}$ = $m_{13}$, thus requiring 
\begin{equation}\label{BimaxmalCond}
f_{[12]} = f_{[13]},
\end{equation}
as can been seen from Eq.(\ref{Eq:MajoranaMass1}). In the case that the heavy-lepton-contributions dominate the charged-lepton-contributions, which will be the case, our estimates of Eq.(\ref{Eq:MajoranaMass2}) yield
\begin{eqnarray}
&&c_0 \approx f\lambda_5m_{E^1}v_\rho^2\left( f^E_{\{12\}}m_{E^2}+ f^E_{\{13\}}m_{E^3} \right)I_{univ},\quad
c_b  \approx f^{\prime E}_{\{11\}}\lambda_5 m_{E^1}^2 v_\rho^2I_{univ}
\label{Eq:MajoranaMass3}
\end{eqnarray}
with $f$ = $f_{[12]}$ = $f_{[13]}$, which in turn give a relation of 
\begin{equation}
\Delta m_\odot^2 \approx \epsilon\frac{\sqrt{2}c_b}{c_0} \Delta m_{atm}^2 \approx \epsilon 
\frac{
\sqrt{2}f^{\prime E}_{\{11\}}m_{E^1}
}
{
f^E_{\{12\}}m_{E^2}+f^E_{\{13\}}m_{E^3}
} \Delta m_{atm}^2,
\label{Eq:m_odod2m_atm2}
\end{equation}
where $\epsilon$ ($\equiv$ $\mu_b/\mu_0$) describes the $U(1)_{L^\prime}$-breaking effects.  The hierarchy of $\Delta m_{atm}^2\gg\Delta m_\odot^2$ is simply ascribed to the breaking of $U(1)_{L^\prime}$ due to $\epsilon$ unless the couplings of $f^E_{\{1i\}}$ ($i$=2,3) and $f^{\prime E}_{\{11\}}$ as well as the masses of $E^1$ and $E^{2,3}$ develop large hierarchy.  In the case that $f^E_{\{1i\}}$ $\sim$ $f^{\prime E}_{\{11\}}$, we reach a simpler relation expressed in terms of heavy lepton masses:
\begin{equation}
\Delta m_\odot^2 \sim \epsilon \frac{\sqrt{2}m_{E^1}}{m_{E^2}+m_{E^3}} \Delta m_{atm}^2.
\label{Eq:Simpler_m_odod2m_atm2}
\end{equation}
Since $\Delta m_{atm}^2$ $\sim$ 3$\times$10$^{-3}$ eV$^2$, we finally obtain $\Delta m_\odot^2$ $\sim$ $\epsilon \Delta m_{atm}^2$ for $m_{E^1}$ $\sim$ $m_{E^{2,3}}$, which, thus, covers the LMA solution for $\epsilon\sim 0.01-0.1$, the LOW solution for $\epsilon$ $\sim$ $10^{-4}$, the QVO solution for $\epsilon$ $\sim$ $10^{-6}$ and the VO solution for $\epsilon$ $\sim$ $10^{-7}$.

More accurate numerical analysis is done by Gaussian integral method to evaluate the integral of Eq.(\ref{Eq:I_1}).  For the various parameters in this model, where the scale of the model is set by $v_\chi$ that breaks $SU(3)_L \times U(1)_N$ down to $SU(2)_L \times U(1)_Y$, we make the following assumptions on relevant free parameters to compute $\Delta m_{atm}^2$ and $\Delta m_\odot^2$ in Eq.(\ref{Eq:DeltaM}): 
\begin{description}
    \item{-} $v_{\eta,\rho} = v_{weak}/\sqrt{2}$ to set $(v^2_\eta+v^2_\rho)^{1/2}$ = $v_{weak}$ for the weak boson masses, where $v_{weak}$ = $( 2{\sqrt 2}G_F)^{-1/2}$ = 174 GeV,
    \item{-} $m_{\eta,\rho} = v_{weak}$ and $m_\xi = 3v_{weak}$ to suppress the $\xi$-contributions in low-energy phenomenology, 
    \item{-} $v_\chi \gg v_{weak}$ to suppress interactions for heavy leptons, exotic quarks and exotic gauge bosons, from which $v_\chi$ $=$ $10v_{weak}$, $m_\chi = v_\chi$ and $m_{E^1} \sim m_{E^2} \sim m_{E^3} \sim ev_\chi$ are taken, where $e$ stands for the electromagnetic coupling,
    \item{-} $m_{k,k^\prime} = v_\chi$ to suppress the $k^{++}$- and $k^{\prime ++}$-contributions in low-energy phenomenology, 
    \item{-} $\lambda_5 = 2\times 10^{-6}$ (with $\lambda_4 \sim \lambda_5$) to reproduce $\Delta m_{atm}^2$ and $\mu_b \leq v_\chi/10$ to cover the LMA solution, where  $\lambda_5 \ll 1$ and $\mu_b \ll v_\chi$ can be considered to be natural because these couplings explicitly break the $L$- or $L^\prime$-conservation \cite{tHooft} while $f_{[1i]}$ $=$ $e$, $f^E_{\{1i\}}$ $=$ $f^{\prime E}_{\{11\}}$ $=$ 1 and $\mu_0 = v_\chi$ because the absence of these couplings does not enhance symmetries of the model. 
\end{description}
The magnitude of $\lambda_5$ $\sim$ $10^{-6}$ can been seen from the approximation of $\Delta m_{atm}^2$ $\sim 3.4\times 10^4 \lambda_5^2 (v_\chi/1~{\rm GeV})^2$ eV$^2$ using $I_{univ}$ $\sim$ $1/[(16\pi^2)^2v^4_\chi$], which is set equal to $\sim 3 \times 10^{-3}$ eV$^2$. Our numerical analysis finally reproduces $\Delta m_{atm}^2$ = $3.03 \times 10^{-3}$ eV$^2$ for $\lambda_5=2\times 10^{-6}$ and leads to $\Delta m_\odot^2$ = ($0.21 - 2.14) \times 10^{-4}$ eV$^2$ as the LMA solution for $\epsilon\sim 0.01-0.1$. 

It is useful to estimate $U_{e3}$, the matrix element that connects $\nu_e$ to $\nu_3$, since it can be used to classify various models for neutrino oscillations\cite{Ue3}. This element of $U_{e3}$ is subject to the experimental constraint of $\vert U_{e3}\vert^2\mapleq 0.015-0.05$ \cite{Chooz}.  In the present mass matrix of Eq.(\ref{Eq:MassMatrix}). this element is calculated to be
\begin{equation}
U_{e3}=-\frac{1}{m_\nu}\left[
\left( \cos^2\vartheta-\sin^2\vartheta\right)m^\prime_{23}+\cos\vartheta\sin\vartheta\left( m^\prime_{33}-m^\prime_{22}\right)
\right],
\label{Eq:Ue3}
\end{equation}
which turns out to vanish since $m^\prime_{ij}\propto f_{[1i]}f_{[1j]}$ ($i$=2,3) as in Eq.(\ref{Eq:MajoranaMass1}) and $\sin\vartheta/\cos\vartheta=f_{[13]}/f_{[12]}$ specific to the present two-loop kinematics.  One can further check that, up to the third order of $m^\prime_{ij}$, $U_{e3}$ is proportional to Eq.(\ref{Eq:Ue3}) as shown in the Appendix so that $U_{e3}$ vanishes up to the third order. Therefore, we expect that $\vert U_{e3}\vert \mapleq (m^\prime_{ij}/m_\nu)^4$, which is about $(\Delta m^2_\odot/\Delta m^2_{atm})^4\mapleq 10^{-4}$.  It should be noted that this property of the vanishing $U_{e3}$ up to the third order does not rely upon Eq.(\ref{BimaxmalCond}) for the baimaximal mixing.

Two comments are in order.  The first one concerns the choice of $L^\prime$ for $k^{\prime ++}$. If $L^\prime$ = 2 instead of  $L^\prime$ = $-2$, $k^{\prime ++}$ couples to the 2nd and 3rd families.  As a result, we reach the mass matrix with the (1,1)-entry and without other entries.  Then, $m^\prime_{11}$ is roughly proportional to $(m_{E^2}+m_{E^3})^2$, which results in $\Delta m_\odot^2$ $\sim$ $\epsilon \Delta m_{atm}^2(m_{E^2}+m_{E^3})/\sqrt{2}m_{E^1}$ estimated to be the similar magnitude to the one by Eq.(\ref{Eq:Simpler_m_odod2m_atm2}) for $m_{E^1}$ $\sim$ $m_{E^{2,3}}$.  Therefore, the same conclusion comes out. The second one is about phenomenological constraints on masses and couplings of $\xi^+$, $k^{++}$ and $k^{\prime ++}$ because these contributions to leptonic processes of $\mu$- and $\tau$-decays, $e^-e^- \rightarrow e^-e^-$ and $\nu_e e^- \rightarrow \nu_e e^-$ would disturb the well-established low-energy phenomenology.  They are given by, 
\begin{enumerate}
\item for $\mu ^ - \to e^ -  e^ -  e^ + $ and $\mu ^ -  \to e^ -   \gamma$ \cite{Mu_E} induced by the collaboration of $k^{++}$ and $k^{\prime ++}$, 
$\kappa f^\ell_{\{12\}} f^{\prime\ell}_{\{11\}}/\bar m_k^2$ $\mapleq$ $1.2 \times 10^{ - 10}$ GeV$^{ - 2}$ from $B(\mu ^ - \to e^ -  e^ -e^ +)$ $<$ $10^{-12}$ \cite{Data} and $\mapleq$
$2.4 \times 10^{ - 8}$ GeV$^{ - 2}$ from $B(\mu ^ -   \to e^ -   \gamma)$ $<$ $1.2 \times 10^{-11}$ \cite{Data}, where ${\bar m_k}$ $\sim$ $m_k$ $\sim$ $m_{k^\prime}$ and $\kappa$ estimated to be about $\mu _b \mu _0/[16\pi ^2\bar m_k^2]$ ($\ll$ 1) reads the suppression due to the approximate $L^\prime$-conservation accompanied by the loop factor of $(16\pi^2)^{-1}$, which arises from the $h^+$-loop for the mixing of $k^{++}$ and $k^{\prime ++}$,
\item for $\tau ^ -   \to\mu^ -  e^ -  e^ + $ and $\tau ^ -  \to\mu^ -    \gamma$ mediated by $k^{++}$ and by $\xi^+$, 
$\left\vert f^\ell_{\{13\}}f^\ell_{\{12\}}/\bar m_k^2\right\vert$ $\mapleq$ 
$2.1 \times 10^{ - 7}$ GeV$^{ - 2}$ from $B(\tau ^ -   \to \mu^ -  e^ -  e^ + )$ $<$ $1.7 \times 10^{-6}$ \cite{Data} and $\mapleq$ $4.2 \times 10^{ - 6}$ GeV$^{ - 2}$ from $B(\tau ^ -   \to \mu^ -   \gamma )$ $<$ $1.1 \times 10^{-6}$ \cite{Data} and 
$\left\vert f_{[13]} f_{[12]}/m_{\xi^+}^2 \right\vert$ $\mapleq$ $4.2 \times 10^{ - 6}$ GeV$^{ - 2}$ also from $B(\tau ^ -   \to \mu^ -   \gamma)$,
\item for $e^ -  e^ -   \to e^ -  e^ -  $ mediated by $k^{\prime ++}$ \cite{E_E}, 
$\left\vert f^{\prime\ell}_{\{11\}}/m_{k^\prime}\right\vert^2$ $\mapleq$ $4.8 \times 10^{ - 5}$ GeV$^{ - 2}$ and,
\item for $\nu_\mu e^- \to \nu_\mu e^-$ mediated by $\xi^+$ \cite{MuDecay}, 
$\left\vert f_{[12]}/m_{\xi^+}\right \vert^2$ $\mapleq$ $1.7 \times 10^{ - 6}$ GeV$^{ - 2}$.
\end{enumerate}
It should be noted that the leading contribution of $\xi^+$ to $\mu^- \to e^-\gamma$, which gives 
the most stringent constraint on $\xi^+$, is forbidden by the $U(1)_{L^\prime}$-invariant coupling 
structure.  Our choice of $f_{[1i]}$ $\sim$ $e$ with $m_{\xi^+}$ $\sim$ $3v_{weak}$ is consistent with the constraints.  As far as these constraints are satisfied, the charged lepton-exchanges in Eq.(\ref{Eq:MajoranaMass2}) turn out to yield well suppressed contributions compared with those from the heavy-lepton exchanges.
\footnote{
The charged-lepton contributions are suppressed as far as $\lambda_4$ $\mapleq$ $\lambda_5 m_{E^1}m_{E^3}v^2_\rho /(m_em_\tau v^2_\chi )$, roughly, leading to $\lambda_4$ $\mapleq$ 0.8.
} 

In summary, 
we have demonstrated that the solar neutrino mixing based on the currently most favorable LMA solution is compatible with the one discussed in the specific class of $SU(3)_L \times U(1)_N$ gauge models that involves $\psi_L=(\nu^i,\ell^i,E^i)_L^T$ as the lepton triplets and  $\xi=(\xi^{++},\overline{\xi}^+,\xi^+)^T$ as the Higgs triplet. Neutrinos acquire Majorana masses through the two-loop radiative mechanism combined with the approximate conservation of $L^\prime$ (=$L_e-L_\mu-L_\tau$). Our two-loop radiative mechanism utilizes $\xi$ that couples to $\psi^1_L\psi^{2,3}_L$ and $k^{++}$ with $L^\prime$ = 0 that couples to $e_R\ell^i_R$ and to $E^1_RE^i_R$ ($i$=2,3) as well as $k^{\prime ++}$ with $L^\prime$ = $-2$ that couples to $e_Re_R$ and to $E^1_RE^1_R$.  As a result, the neutrino mass matrix exhibits the bimaximal structure if the approximate equality between $f_{[12]}$ and $f_{[13]}$ is realized. The relation of $\Delta m_\odot^2$ $\sim$ $\sqrt{2}\epsilon \Delta m_{atm}^2m_{E^1}/(m_{E^2}+m_{E^3})$ successfully covers the LMA solution.

\bigskip
\noindent
The work of M.Y. is supported by the Grant-in-Aid for Scientific Research No 12047223 from the Ministry of Education, Science, Sports and Culture, Japan.

\bigskip
\bigskip

\centerline{\bf Appendix}

The matrix element $U_{e3}$ can be defined by
\begin{equation}
U_{e3}=\frac{1}{\sqrt{2}}
\left(
c^{(1)}_{31}+c^{(2)}_{31}+c^{(3)}_{31}-c^{(1)}_{32}-c^{(2)}_{32}-c^{(3)}_{32}
\right),
\label{Eq:Appendix_Ue3}
\end{equation}
where $c^{(1,2,3)}_{3i}$ ($i$=1,2) are the coefficients in the perturbative expansion of $\vert \nu_3 \rangle$ as the eigenstate of $\nu_3$ for $M_\nu$ of Eq.(\ref{Eq:MassMatrix}) in terms of $m^\prime_{ij}$ ($i,j$=2,3). In fact, $U_{e3}$ is expressed in terms of unperturbed state of $\vert i^{(0)}\rangle$ ($i$=1,2,3):
\begin{eqnarray}
&&\vert \nu_3 \rangle=
    \left(
    \begin{array}{ccc}
         U_{e3} \\
         \ast  \\
         \ast
    \end{array}
    \right)
= \vert 3^{(0)}\rangle+\sum_{i=1,2,3}\left( 
c^{(1)}_{3i}+c^{(2)}_{3i}+c^{(3)}_{3i}
\right)
\vert i^{(0)}\rangle,
\label{Eq:Appendix_c123}
\end{eqnarray}
up to the third order.  Since $\vert 3^{(0)}\rangle$ has no relevant entry for $U_{e3}$, $\vert 3^{(0)}\rangle$ and $c^{(1,2,3)}_{33}$ can be neglected.  The unperturbed states are defined by $\vert 1^{(0)} \rangle = (1, c, s)^T/\sqrt{2}$, $\vert 2^{(0)} \rangle=(-1, c, s)^T/\sqrt{2}$ and $\vert 3^{(0)} \rangle = (0, -s, c)^T$ for $c=\cos\vartheta$ and $s=\sin\vartheta$, whose masses are given by ($m^{(0)}_1$, $m^{(0)}_2$, $m^{(0)}_3$) = ($m_\nu$, $-m_\nu$, 0), respectively.  

The coefficients can be determined by the well-known formula:
\begin{eqnarray}
&&c^{(1)}_{ij} = -\langle j^{(0)}\vert M^\prime_\nu \vert i^{(0)}\rangle/\left( m^{(0)}_j-m^{(0)}_i\right),
\quad
c^{(2)}_{ij} = -\left(\sum_k c^{(1)}_{ik}\langle j^{(0)}\vert M^\prime_\nu \vert k^{(0)}\rangle-m^{(1)}_ic^{(1)}_{ij}\right) /\left( m^{(0)}_j-m^{(0)}_i\right),
\nonumber \\
&&c^{(3)}_{ij} = -\left(\sum_k c^{(2)}_{ik}\langle j^{(0)}\vert M^\prime_\nu \vert k^{(0)}\rangle-m^{(1)}_ic^{(2)}_{ij}-m^{(2)}_ic^{(1)}_{ij}\right) /\left( m^{(0)}_j-m^{(0)}_i\right),
\nonumber \\
&&c^{(1)}_{ii}=0,
\quad
c^{(2)}_{ii}=-\sum_j\vert c^{(1)}_{ij}\vert^2/2,
\quad
c^{(3)}_{ii}=-\sum_j\left( c^{(2)\ast}_{ij}c^{(1)}_{ij}+c^{(1)\ast}_{ij}c^{(2)}_{ij}\right)/2,
\label{Eq:Appendix_Coeff}
\end{eqnarray}
where
\begin{eqnarray}
&& m^{(1)}_i=\langle i^{(0)}\vert M^\prime_\nu \vert i^{(0)}\rangle,
\quad
m^{(2)}_i=\sum_j c^{(1)}_{ij}\langle i^{(0)}\vert M^\prime_\nu \vert j^{(0)}\rangle,
\label{Eq:Appendix_Masses}
\end{eqnarray}
and $M^\prime_\nu$ is the $3\times 3$ matrix only with $m^\prime_{ij}$ in the same entries as in $M_\nu$.  These equations lead to
\begin{eqnarray}
 U_{e3}&=&\sqrt{2}c^{(1)}_{31}
-
\sqrt{2}
\left[
\left(
\langle 1^{(0)}\vert M^\prime_\nu \vert 2^{(0)}\rangle
+m^{(1)}_1-m^{(1)}_3
\right)
c^{(2)}_{31}
+
\langle 1^{(0)}\vert M^\prime_\nu \vert 3^{(0)}\rangle c^{(2)}_{33}
\right]/m_\nu
\label{Eq:Appendix_Ue3_Result}
\end{eqnarray}
with
\begin{eqnarray}
c^{(1)}_{31} &=& -
\left[
\left( c^2-s^2\right) m^\prime_{23}+cs\left( m^\prime_{33}-m^\prime_{22}\right)
\right]/\sqrt{2}m_\nu,
\quad
c^{(2)}_{33} =-c^{(1)2}_{31},
\nonumber \\
c^{(2)}_{31} &=&
\left(
s^2m^\prime_{22}-2csm^\prime_{23}+c^2m^\prime_{33}
\right)c^{(1)}_{31}
/m_\nu,
\label{Eq:Appendix_c}
\end{eqnarray}
where we have used various relations such as $c^{(1)}_{31}=-c^{(1)}_{32}$, $c^{(2)}_{31}=c^{(2)}_{32}$, $m^{(1)}_1 = m^{(1)}_2$ and so on.  Especially, the relation of $c^{(2)}_{31}=c^{(2)}_{32}$ makes the second order $U_{e3}$ to vanish.  We, then, find that $U_{e3}$ is proportional to $c^{(1)}_{31}$, which vanishes because of the proportionality of $m^\prime_{ij}\propto f_{[1i]}f_{[1j]}$ ($i,j$=2,3) indicated by Eq.(\ref{Eq:MajoranaMass1}) with $f_{[12]} \propto c$ and $f_{[13]} \propto s$.



\noindent
\centerline{\bf Table Captions}
\\
\\
\centerline{TABLE \ref{Tab:Lnumber}: $L$ and $L^\prime$ quantum numbers.}
\\
\\
\noindent
\centerline{\bf Figure Captions}
\begin{figure}
\caption{Two-loop diagrams via (a) $L^\prime$-conserving interactions and via (b) $L^\prime$-violating interactions.}
\label{Fig:loopDiagrams}
\end{figure}


\begin{table}[ht]
    \caption{\label{Tab:Lnumber}$L$ and $L^\prime$ quantum numbers.}
    \begin{center}
    \begin{tabular}{ccccccc}
    \hline
        Fields  & $\eta,\rho,\chi$ & $\xi$ 
				& $k^{++}$ & $k^{\prime ++}$
                & $\psi_L^1,\ell_R^1,E_R^1$ & $\psi_L^{2,3},\ell_R^{2,3},E_R^{2,3}$ \\ \hline
        $L$     & 0                & $-$2   
                & $-$2      & $-$2 
                & 1                         &  1\\ \hline
    $L^\prime$  & 0                &  0 
                &  0               & $-$2 
                & 1                         & $-$1 \\ \hline
    \end{tabular}
    \end{center}
\end{table}
\centerline{\epsfbox{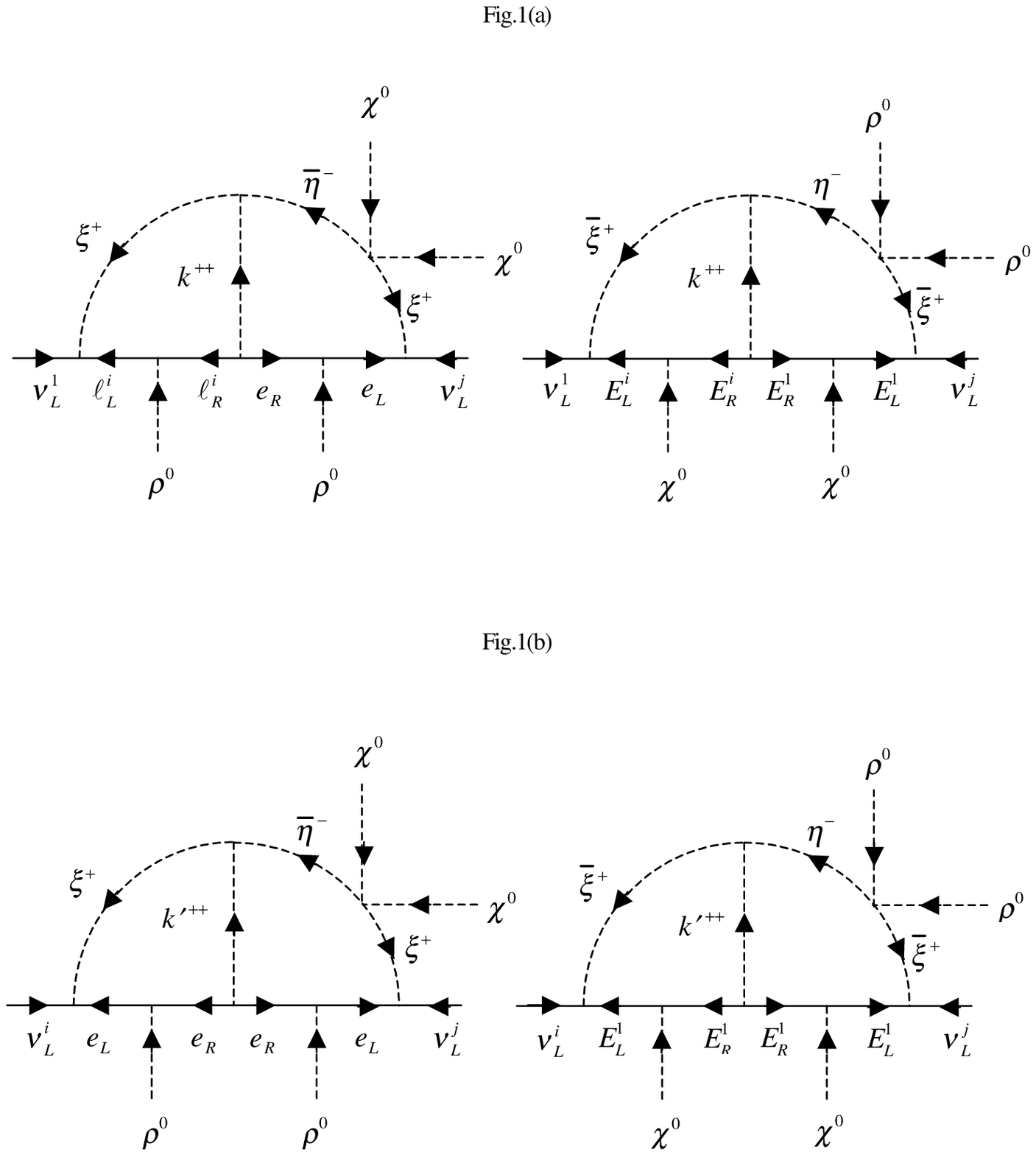}}

\end{document}